\documentclass[aps, pra, floatfix]{revtex4}
\usepackage{graphicx}
\usepackage{amsmath}
\usepackage{setspace}

\begin{document}
\title{Insights into the semiclassical Wigner treatment of bimolecular collisions}

\author{L. Bonnet{\footnote{Email: laurent.bonnet@u-bordeaux1.fr}}}

\affiliation{CNRS, Univ. Bordeaux, ISM, UMR 5255, 33405, Talence, France}

\date{\today}

\begin{abstract}
\noindent The semiclassical Wigner treatment of bimolecular collisions, proposed by Lee and Scully 
on a partly intuitive basis [J. Chem. Phys. 73, 2238 (1980)], is derived here from first principles. 
The derivation combines E. J. Heller's ideas [J. Chem. Phys. 62, 1544 (1975); 65, 1289 (1976); 75, 186 (1981)], 
the backward picture of molecular collisions [L. Bonnet, J. Chem. Phys. 133, 174108 (2010)] 
and the microreversibility principle. 
\end{abstract}
 
\maketitle

\section{Introduction}
\label{I}

Quantum mechanical (QM) calculations of molecular reaction dynamics
\cite{Nyman,Maha,Launay,Althorpe,Schatz,Bruno,Han,Dario,Aron,Manthe,Xiao}
are generally much heavier than quasi-classical trajectory calculations \cite{PR,ST,Aoiz}, 
in particular for polyatomic processes \cite{Joa,Miguel,Cz1,Cz2}. Hence, for several decades, intense
researches aim at building semiclassical methods taking into account the largest quantum effects while keeping with the classical description of nuclear motions \cite{SC1,SC2,SC3,SC4,SC5,SC5a,SC6a,SC6,SC7,SC8,SC9,SC9a,SC10,SC11,Arbelon}. 
Moreover, developping such methods naturally leads to shed light on the complex frontier between the quantum and classical descriptions of motion. Last but not least, semiclassical approaches may be used as powerful interpretative tools 
to rationalize the dynamics of molecular processes. 

\begin{figure}
  \includegraphics[width=200mm]{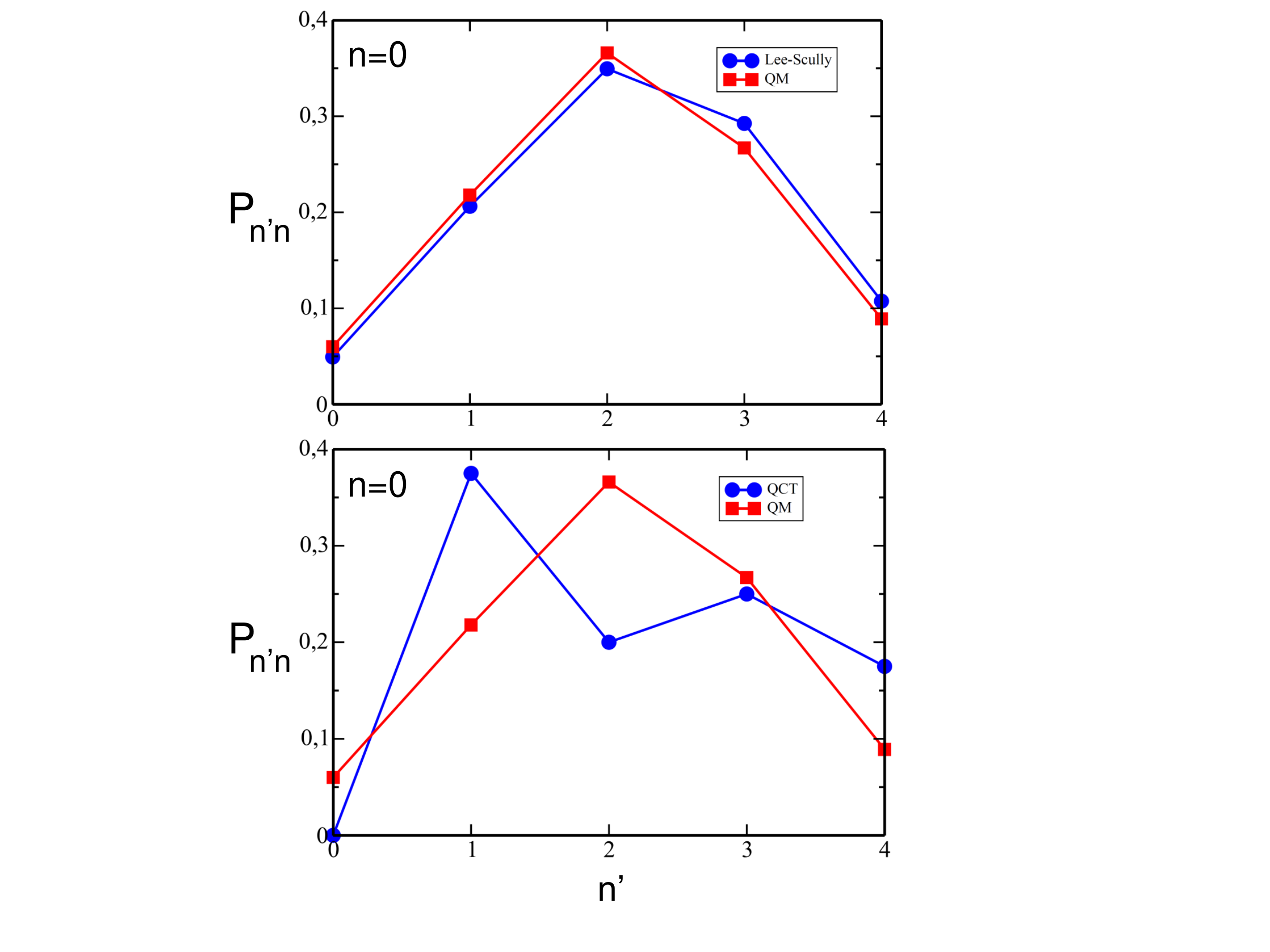}
  \caption{Final vibrational state distribution for the Secrest-Johnson model
  of collinear collision He+H$_2(n)$ $\longrightarrow$ He+H$_2(n')$ at a total energy of 10 and for $n=0$. 
  Lee-Scully predictions, given by Eq.~\eqref{2B11}, and
  QCT ones are compared with QM results in the upper and lower panels, respectively. QM and QCT data come from
  refs.~\cite{Secrest} and \cite{SC5a}, respectively. 
   \label{fig:1}}
\end{figure}

\begin{figure}
  \includegraphics[width=180mm]{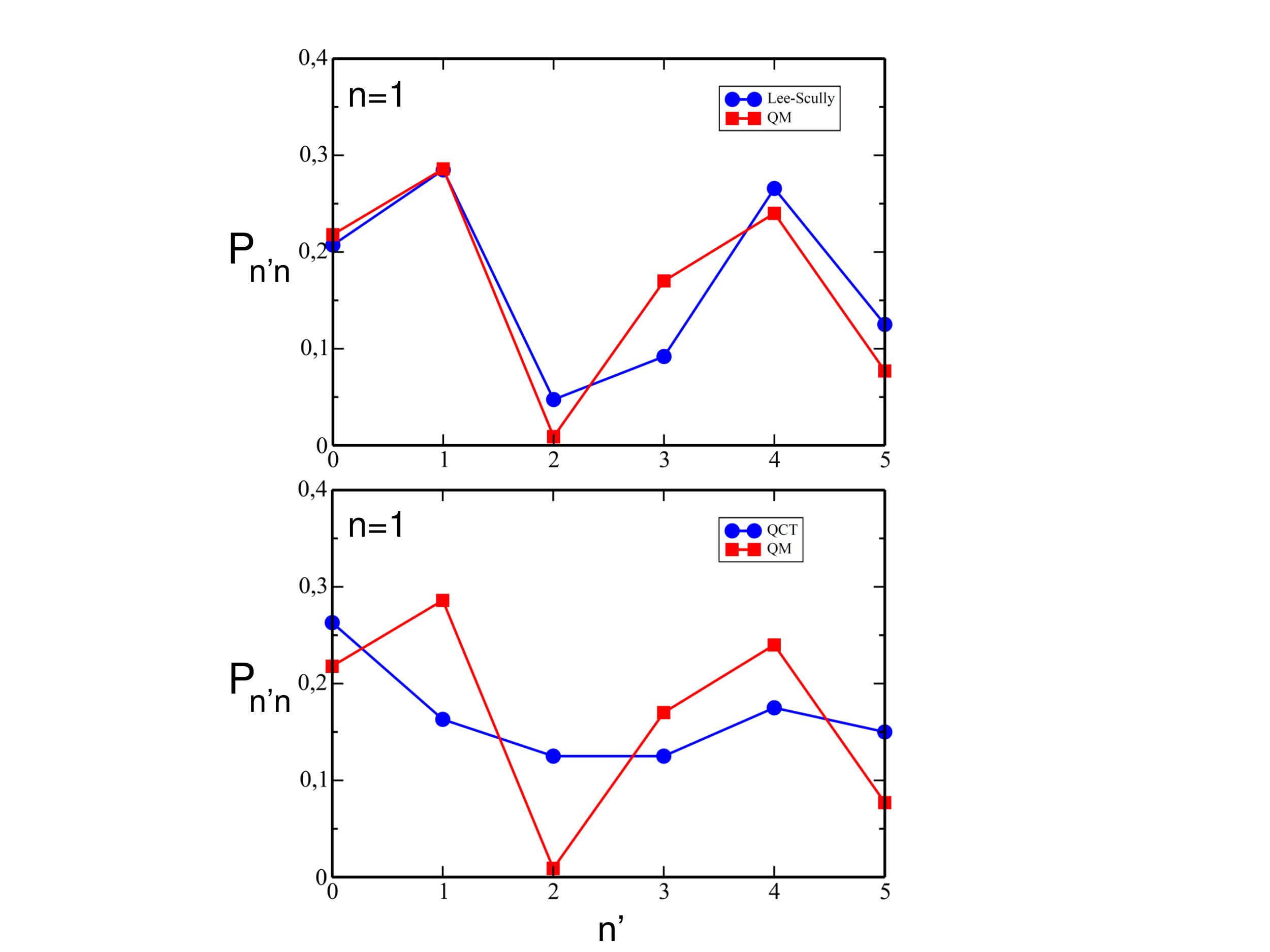}
  \caption{Same as Fig.~\ref{fig:1} for $n=1$.  
   \label{fig:2}}
\end{figure}

The semiclassical Wigner treatment of bimolecular collisions proposed by Lee and Scully is among them \cite{SC5a}.
When applied to the collinear inelastic collision between He and H$_2$, or He and HBr, which involve strong quantum interferences and/or classically forbidden vibrational transitions, this approach leads to final state
populations in very good agreement with exact quantum ones (at least for the lowest vibrational states
of the initial diatom), contrary to the quasi-classical trajectory (QCT) method \cite{SC5a}. 
This is illustrated in Figs.~\ref{fig:1} and~\ref{fig:2} for He plus H$_2$ (note that these results are
particularly relevant as comparative semiclassical/quantum studies of
collinear processes usually provide more stringent tests of the validity of semiclassical approaches
than similar studies for realistic three-dimensional collisions). In addition, the treatment 
of Lee and Scully is particularly simple to apply.
Hence, it is potentially interesting to study realistic molecular collisions involving certain quantum effects (other
than tunneling though a potential energy barrier).

However, the original derivation of this treatment \cite{SC5a} appears to rest on somewhat arbitrary basis, as shown 
later below. The goal of the present work is thus to put this treatment on firmer theoretical grounds by deriving 
it from first principles. 

This work is mainly motivated
by the recent extention of the semiclassical Wigner treatment of Heller \cite{SC6a,SC6} to triatomic (or triatomic-like polyatomic) photodissociations \cite{Arbelon}. 
For the first time, rotational motions were taken into account in the theory, thus making
it applicable to realistic fragmentations, and comparison between its predictions and rigorous quantum results in the case of Guo's triatomic-like model of methyl iodide photodissociation \cite{Guo} showed nearly quantitative agreement. 
Since the method of Lee and Scully
bears strong resemblance with the one of Brown and Heller, we hope to be able in the future to include rotational motions 
in the former so as to make it applicable to realistic bimolecular processes. The present work is a preliminary 
step towards this goal.

The inelastic collision is defined in section \ref{II}. The Lee-Scully method is summarized in section \ref{III}. Its derivation from first principles is presented in section \ref{IV}. Section \ref{V} concludes.

\section{Collisional system}
\label{II}

We consider the collinear inelastic collision between atom A and diatom BC($n$).
The process takes place in the electronic ground state. $R$ is the distance
between A and the center-of-mass of BC, and $r$ is the BC bond length. $P$ and $p$ are the momenta conjugate
to $R$ and $r$, respectively. The classical function of Hamilton reads
\\
\begin{equation}
H = \frac{P^2}{2\mu}+\frac{p^2}{2m}+V(R,r).
\label{1}
\end{equation}
\\
$\mu$ is the reduced mass of A with respect to BC and $m$, the one of BC. $V(R,r)$
is the potential energy of ABC associated with the electronic ground state. 
In the asymptotic channel, BC is assumed to be a harmonic oscillator of frequency $\omega/2\pi$.
Its initial vibrational state, of energy 
\\
\begin{equation}
E_{n} = \hbar \omega (n+1/2),
\label{1c}
\end{equation}
\\
is denoted $\chi_{n}(r)$. 
This state is real and normalized to unity. 
The total energy available to the separated fragments is denoted $E$. The initial collision energy is thus $E_c=E-E_{n}$
and the initial value of $P$ is 
\\
\begin{equation}
P_i=-[2\mu E_c]^{1/2}.
 \label{1a}
\end{equation}
\\
The Hamiltonian operator is denoted $\hat{H}$.

The numerical results presented throughout this work were obtained within the framework of the Secrest-Johnson model 
of He+H$_2$ collision \cite{Secrest}. In this model, the potential energy is approximated by
\\
\begin{equation}
V(R,r)=\frac{1}{2} m \omega^2(r-r_e)^2+exp[\alpha (r-r_e-R)].
\label{1b}
\end{equation} 
\\
$\mu$ and $m$ were kept at 2/3 and 1, respectively, $\hbar$ at 1, $\omega$ at 1, $\alpha$ at 0.3 and $E$ at 10, otherwise
stated. The value
of the equilibrium bond length $r_e$ can be arbitrarily chosen. 

The central quantity of this work, already shown for He+H$_2$ in Figs.~\ref{fig:1} and~\ref{fig:2}, is the probability $P_{n'n}$ of transition from BC($n$) to BC($n'$).

\section{Lee-Scully method}
\label{III}

Lee and Scully consider a hybrid model where the initial and final states of BC are treated 
quantum mechanically, while the initial and final translational states as well as the whole dynamics within 
the interaction region are treated classical mechanically. This choice makes sense, in particular
to QCT users accustomed to introduce \emph{ad-hoc} corrections in their trajectory calculations, but from
a rigorous point of view, it is arbitrary. 

These assumptions lead to the following developments. The final vibrational state of BC can be written as
\\
\begin{equation}
\phi_f(r) = \sum_{j=0}^{\infty} a_j \; \chi_{j}(r),
\label{2B1}
\end{equation}
\\
for the $\chi_{j}(r)$'s form a complete basis. Hence, $P_{n'n}$ satisfies
\\
\begin{equation}
P_{n'n} = |a_{n'}|^2 = \left|\int\;dr \; \chi_{n'}{}^*(r) \; \phi_f(r)\right|^2.
\label{2B2}
\end{equation}
\\
Now, any quantity $|Q|^2$ defined by 
\begin{equation}
|Q|^2 = \left|\int\;dr \; \Psi_1{}^*(r) \; \Psi_2(r)\right|^2
\label{2B3}
\end{equation}
\\
is also rigorously given by the phase space overlap
\\
\begin{equation}
|Q|^2 = 2\pi \hbar \int\;dr dp \; \rho_1(r,p) \; \rho_2(r,p),
\label{2B4}
\end{equation}
\\
where $\rho_l(r,p)$, $l=1$ or 2, is the Wigner density defined by
\\
\begin{equation}
\rho_l(r,p) = \frac{1}{\pi \hbar}\int\;ds \; e^{2ips/\hbar}\;\Psi_l{}^*(r+s)\;\Psi_l(r-s)
\label{2B5}
\end{equation}
\\
\cite{SC5a,SC6a,SC6,SC10,Arbelon,Wig,LS}. 
A pedestrian demonstration of the strict equivalence between Eq.~\eqref{2B3} and Eqs.~\eqref{2B4} and \eqref{2B5} 
is given in Appendix B of ref.~\cite{SC10}. Consequently, $P_{n'n}$ can be rewritten as
\\
\begin{equation}
P_{n'n} = 2\pi \hbar \int\;dr dp \; \rho^{vib}_{n'}(r,p) \; \rho_f(r,p),
\label{2B6}
\end{equation}
\\
where $\rho^{vib}_{n'}(r,p)$ and $\rho_f(r,p)$ are the Wigner distributions corresponding to 
$\chi_{n'}(r)$ and $\phi_f(r)$, respectively.

From the well known analytical expression of $\chi_{n'}(r)$ and Eq.~\eqref{2B5}, 
$\rho^{vib}_{n'}(r,p)$ can be shown to satisfy \cite{SC6}
\\
\begin{equation}
 \rho^{vib}_{n'}(r,p) = \frac{(-1)^{n'}}{\pi\hbar}L_{n'}(\xi)\exp(-\xi/2),
 \label{2B7}
\end{equation}
\\
where $L_{n'}(\xi)$ is the ${n'}$th Laguerre polynomial,
\\
\begin{equation}
 \xi = \frac{4E_v}{\hbar\omega}
 \label{2B8}
\end{equation}
and
\begin{equation}
 E_v = \frac{p^2}{2m} + \frac{1}{2}m\omega^2(r-r_e)^2.
 \label{2B8a}
\end{equation}
\\
$E_v$ appears to be the BC vibrational energy.

$\rho_f(r,p)$ is estimated as follows. $\phi_f(r)$ is supposed to result from the ``propagation'' in time of
$\chi_{n}(r)$ from the reagents onto the products.
This loose statement leads, however,
 to a practical method when extended to the Wigner densities $\rho_f(r,p)$ and $\rho^{vib}_{n}(r,p)$.
$\rho^{vib}_{n}(r,p)$ can indeed be rewritten as
\\
\begin{equation}
 \rho^{vib}_{n}(r,p) = \int\;dr_i dp_i \; \rho^{vib}_{n}(r_i,p_i) \delta(r-r_i) \delta(p-p_i).
 \label{2B9}
\end{equation}
\\
Propagating $\rho^{vib}_{n}(r,p)$ in time amounts to move each delta peak $\delta(r-r_i) \delta(p-p_i)$ in the $(r,p)$ plane along the classical path starting from $(r_i,p_i)$. This path is also specified by $R=R_i$, supposed to be large enough for 
BC to vibrate freely, and $P = P_i$. 
Once the trajectories are back to $R=R_i$, where BC has reached its final vibrational state, each delta peak is
located at a given point of coordinates $(r_f,p_f)$, both depending implicitely on $(r_i,p_i)$. Consequently,
\\
\begin{equation}
 \rho_{f}(r,p) = \int\;dr_i dp_i \; \rho^{vib}_{n}(r_i,p_i) \delta(r-r_f) \delta(p-p_f).
 \label{2B10}
\end{equation}
\\
Replacing $\rho_{f}(r,p)$ in Eq.~\eqref{2B6} by the right-hand-side (RHS) of the above equation and integrating
over $r$ and $p$ finally leads to the Lee-Scully expression 
\\
\begin{equation}
P_{n'n} = 2\pi \hbar \int\;dr_i dp_i \; \rho^{vib}_{n'}(r_f,p_f) \; \rho^{vib}_n(r_i,p_i)
\label{2B11}
\end{equation}
\\
(see the third identity of Eq. (21) in ref. \cite{SC5a}). The Lee-Scully results presented in Figs.~\ref{fig:1} and \ref{fig:2} have been obtained from a simple Monte-Carlo method based on 10$^6$ trajectories, with $r_i$ and $p_i$ both randomly selected within the range [-6,6], and $R_i$ taken at 20.

\section{Derivation of the Lee-Scully method from first principles}
\label{IV}

We now derive a formulation which closely parallels the time-dependent quantum description as far as possible.

\subsection{Link between transition amplitudes and time-evolved wave-packet}
\label{IV.A}

Only the main lines of the derivation of transition amplitudes in terms of a time-evolved wave-packet are given 
in this section, for it is a known result \cite{SC3}. For clarity's sake, however, more details are given in Appendix A.

We consider at time 0 the wave-packet defined by:
\\
\begin{equation}
\Psi_0(R,r) = \frac{1}{(2\pi)^{1/2}}\int\;dk\;g(k)e^{-ik(R-R_i)}\chi_n(r)
\label{2}
\end{equation}
with
\begin{equation}
g(k) = \left[\frac{1}{\pi^{1/2}\epsilon}\right]^{1/2}e^{-\frac{1}{2}\left[(k-k_i)/\epsilon\right]^2}.
\label{3}
\end{equation}
\\
$\Psi_0(R,r)$ is normalized to unity. Here,
$R_i$ is sufficiently large for the whole wave-packet to lie within the asymptotic region where $V(R,r)$ does not
depend on $R$ and the vibrational and 
translational motions are uncoupled. $k_i$ has a given positive value, thus making the wave-packet moving inward 
for sufficiently small values of $t$ (propagation is performed forward in time, otherwise stated). 
The spreading of the wave-packet is inversely proportional to $\epsilon$. 

Let $\phi_E^n(R,r)$ be the state of inelastic scattering between A and BC at energy $E$, with unit
incoming flux in channel $n$. In the asymptotic region, $\phi_E^{n}(R,r)$ can be written as
\\
\begin{equation}
\phi_E^n(R,r) = \left[\frac{\mu}{2\pi\hbar^2 k_n}\right]^{1/2}e^{-ik_n R}\chi_n(r)
+\sum_{n"}\;S_{n"n}\;\left[\frac{\mu}{2\pi\hbar^2 k_{n"}}\right]^{1/2}e^{ik_{n"}R}\chi_{n"}(r)
\label{4}
\end{equation}
with 
\begin{equation}
k_{n"}=\frac{1}{\hbar}[2\mu(E-E_{n"})]^{1/2}.
\label{5}
\end{equation}
\\
The $S_{n"n}$'s are transition amplitudes, or \emph{S}-matrix elements. 
The projection of $\phi_E^n(R,r)$ on $\Psi_0(R,r)$ is given by
\\
\begin{equation}
C_n(E) = \int\;dRdr\;{\phi_E^n}{}{}^*(R,r)\Psi_0(R,r).
\label{6}
\end{equation}
\\
The time-evolved wave-packet can then be written as
\\
\begin{equation}
\Psi_t(R,r) = \sum_j\;\int\;dE'\;C_j(E'){\phi_{E'}^n}(R,r)e^{-iE't/\hbar}.
\label{7}
\end{equation}
\\
We now consider at an infinite time the projection
\\
\begin{equation}
Q_{n'n} = \lim_{t \to +\infty} 
\int\;dRdr\;\left[\frac{\mu}{2\pi\hbar^2 k_{n'}}\right]^{1/2}e^{-ik_{n'}R}\chi_{n'}(r)\Psi_t(R,r)
\label{8}
\end{equation}
\\
of the outgoing free state associated with $E$ and $n'$ onto the time-evolved wave-packet.
From Eqs.~\eqref{4} and \eqref{6}-\eqref{8}, it is shown in Appendix A that 
\\
\begin{equation}
g(k_n)e^{ik_n R_i} S_{n'n}+g(-k_n)e^{-ik_n R_i} \delta_{n'n} = 
\left[\frac{\hbar^2 k_n}{\mu}\right]^{1/2} Q_{n'n}\;e^{iEt/\hbar}.
\label{9}
\end{equation}
\\
This is an important expression of the time-dependent approach to semiclassical dynamics
derived by Heller in the mid-seventies \cite{SC3}. Eq.~\eqref{9} is analogous to Eq.~(4.4) in Heller's work.
The only difference is that in the present work, $g(k)$ (see Eq.~\eqref{3}) can be non zero for negative values of $k$.
This implies the Kronecker symbol $\delta_{n'n}$ in Eq.~\eqref{9}, not present in Eq.~(4.4) of ref.~\cite{SC3}.

\subsection{Expressing the projection $Q_{n'n}$ in terms of Wigner densities}
\label{IV.B}

Analogously to what we have seen in section \ref{III}, any quantity $|Q|^2$ defined by 
\\
\begin{equation}
|Q|^2 = \left|\int\;dR dr \; \Psi_1{}^*(R,r) \; \Psi_2(R,r)\right|^2
\label{BB.1}
\end{equation}
can be rewritten as
\begin{equation}
|Q|^2 = (2\pi \hbar)^2\int\;dR dr dP dp \; \rho_1(R,r,P,p) \; \rho_2(R,r,P,p)
\label{BB.2}
\end{equation}
\\
where $\rho_l(R,r,P,p)$, $l=1$ or 2, is the Wigner density defined by
\\
\begin{equation}
\rho_l(R,r,P,p) = \frac{1}{(\pi \hbar)^2}\int\;ds_R ds_r \; 
e^{2i(Ps_R+ps_r)/\hbar}\;
\Psi_l{}^*(R+s_R,r+s_r)\;
\Psi_l(R-s_R,r-s_r)
\label{BB.3}
\end{equation}
\\
\cite{SC6a,SC6,SC10,Arbelon}. The strict equivalence between Eq.~\eqref{BB.1} and Eqs.~\eqref{BB.2} and \eqref{BB.3} 
can be proved by following the reasoning presented in Appendix B of ref.~\cite{SC10} limited, however, to the case of one configuration space coordinate. In the present case of two spatial coordinates, the developments are more tedious, but present no difficulty. 

Setting
\begin{equation}
\Psi_1(R,r) = \left[\frac{\mu}{2\pi\hbar^2 k_{n'}}\right]^{1/2}e^{ik_{n'}R}\chi_{n'}(r)
\label{BB.4}
\end{equation}
and
\begin{equation}
\Psi_2(R,r) = \Psi_t(R,r),
\label{BB.5}
\end{equation}
\\
we arrive from Eqs.~\eqref{BB.2} and \eqref{BB.3} at  
\\
\begin{equation}
|Q_{n'n}|^2 = (2\pi \hbar)^2
\lim\limits_{t \to +\infty} 
\int\;dR dr dP dp\; \rho_t(R,r,P,p)
\; \rho^{tr}_{n'}(R,P)\; \rho^{vib}_{n'}(r,p),
\label{BB.6}
\end{equation}
\\
where $\rho_t(R,r,P,p)$ is related to $\Psi_t(R,r)$ by Eq.~\eqref{BB.3},
the translational Wigner distribution $\rho^{tr}_{n'}(R,P)$ reads
\\
\begin{equation}
\rho^{tr}_{n'}(R,P) = \frac{1}{\pi \hbar}\int\;ds \; 
e^{2iPs/\hbar}\;
\left[\frac{\mu}{2\pi\hbar^2 k_{n'}}\right]\;
e^{-ik_{n'}(R+s)}\;
e^{ik_{n'}(R-s)}
\label{BB.7}
\end{equation}
\\
and $\rho^{vib}_{n'}(r,p)$ is given by Eqs.~\eqref{2B7}-\eqref{2B8a}
\\
Using Eq.~\eqref{A2}, Eq.~\eqref{BB.7} transforms to
\\
\begin{equation}
\rho^{tr}_{n'}(R,P) = \left[\frac{\mu}{2\pi\hbar^2 k_{n'}}\right]\;
\frac{2}{\hbar}\ \delta[2(P-\hbar k_{n'})/\hbar]
\label{BB.10}
\end{equation}
or equivalently, 
\begin{equation}
\rho^{tr}_{n'}(R,P) = \frac{1}{2\pi\hbar} \delta\left[\frac{P^2}{2\mu}-\frac{\hbar^2 k_{n'}^2}{2\mu}\right]\Theta(P),
\label{BB.11}
\end{equation}
\\
$\Theta$ being the function of Heaviside. Eq.~\eqref{BB.10} is indeed readily obtained from Eq.~\eqref{BB.11} by means of theorem~\eqref{A7}.

\subsection{Semiclassical approximation of $|Q_{n'n}|^2$}
\label{IV.C}

Eq.~\eqref{BB.6} provides a formally exact quantum expression of $|Q_{n'n}|^2$. 
We now introduce in the formulation the following classical ingredient: 
we consider $\rho_{\tau}(R,r,P,p)$ as a solution of the Liouville equation \cite{Gold} instead of the Wigner one 
\cite{SC6a,Wig,LS}, i.e., we classically propagate it from $\tau=0$ to $\tau=t$. 
In the framework of this assumption, we have
\\
\begin{equation}
\rho_{t}(R,r,P,p)dRdrdPdp = \rho_{0}(R_0,r_0,P_0,p_0)dR_0dr_0dP_0dp_0.
\label{20}
\end{equation}
\\
In this identity, $(R,r,P,p)$ should be understood as the dynamical state of ABC reached at time $t$ when starting from 
the initial state $(R_0,r_0,P_0,p_0)$ at time 0. Eq.~\eqref{20} expresses the fact that the probability to lie within $dRdrdPdp$
does not depend on $t$, a property due to the deterministic nature of classical mechanics.

Using Eq.~\eqref{20}, on may rewrite Eq.~\eqref{BB.6} as
\\
\begin{equation}
|Q_{n'n}|^2 = (2\pi \hbar)^2
\lim\limits_{t \to +\infty} 
\int\;dR_0dr_0dP_0dp_0\; \rho_{0}(R_0,r_0,P_0,p_0)
\; \rho^{tr}_{n'}(R_t,P_t)\; \rho^{vib}_{n'}(r_t,p_t),
\label{21}
\end{equation}
\\
where for clarity's sake, $(R,r,P,p)$ in $\rho^{tr}_{n'}(R,P) \rho^{vib}_{n'}(r,p)$ 
have been denoted $(R_t,r_t,P_t,p_t)$ in order 
to emphasize that these coordinates determine the dynamical state of ABC at time $t$.

From Eqs.~\eqref{2},~\eqref{3} and~\eqref{BB.3}, one may show after some mathematical steps
presenting no particular difficulty that
\\
\begin{equation}
\rho_{0}(R_0,r_0,P_0,p_0) = \frac{1}{\pi\hbar} e^{-\epsilon^2(R_0-R_i)^2}
e^{-\left(\frac{P_0+\hbar k_i}{\hbar \epsilon}\right)^2} \rho^{vib}_{n}(r_0,p_0),
\label{22}
\end{equation}
\\
with $\rho^{vib}_{n}(r_0,p_0)$ given by Eqs.~\eqref{2B7}-\eqref{2B8a}. 

In fact, $\rho^{vib}_{n'}(r_t,p_t)$ depends on the vibrational energy at $t$ (see Eqs.~\eqref{2B7}-\eqref{2B8a}).
Moreover, this energy is a constant of motion beyond the frontier between the interaction region and the free products,
defined by a given value $R_f$ of $R$. Therefore, 
$\rho^{vib}_{n'}(r_t, p_t)$ can be replaced by $\rho^{vib}_{n'}(r_f, p_f)$ in Eq.~\eqref{21}, where $r_f$ and $p_f$
are, respectively, the values of $r_t$ and $p_t$ at the previous frontier. For exactly the same reason, $P_t$ can be replaced by 
$P_f$ in Eq.~\eqref{21}. We thus have
\\
\begin{equation}
|Q_{n'n}|^2 = (2\pi \hbar)^2 
\int\;dR_0dr_0dP_0dp_0\; \rho_{0}(R_0,r_0,P_0,p_0)
\; \rho^{tr}_{n'}(R_f,P_f)\; \rho^{vib}_{n'}(r_f,p_f).
\label{23}
\end{equation}

\subsection{Validity condition of the semiclassical approximation}
\label{IV.D}

The validity of Eq.~\eqref{23} is conditioned by the ability of classical mechanics to propagate 
$\rho_{t}(R,r,P,p)$ in a realistic way, i.e., such as it stays sufficiently close to the exact time-evolved Wigner density, solution of the Wigner equation \cite{SC6a,Wig,LS}, up to the products. 

\begin{figure}
  \includegraphics[width=180mm]{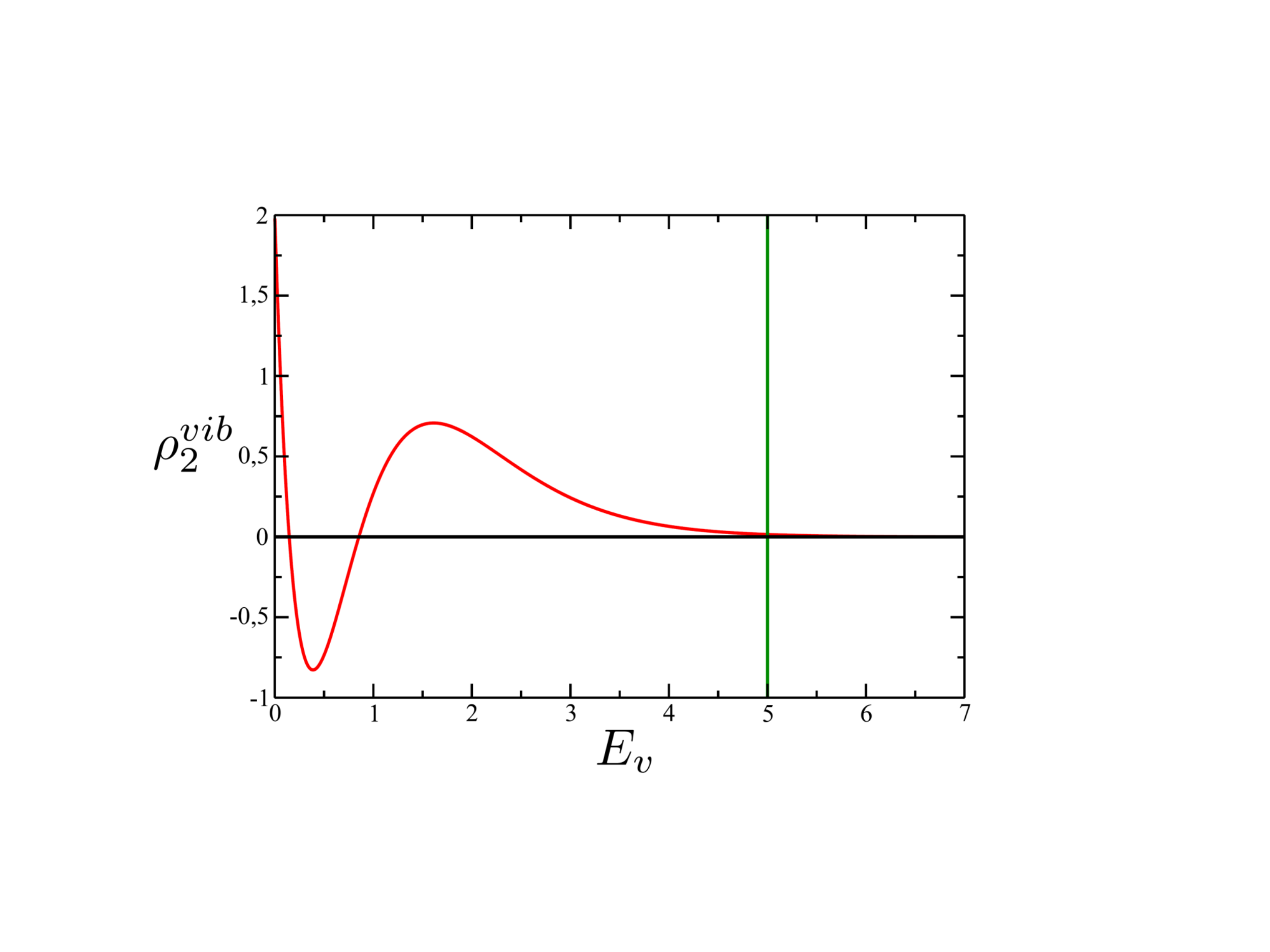}
  \caption{Wigner distribution $\rho^{vib}_2$ for the harmonic oscillator (see Eqs.~\eqref{2B7}-\eqref{2B8a}) in terms of the    
   vibrational energy $E_v$ (red curve). On the RHS of the green vertical line, defined by $E_v = E_2+2.5=5$ 
   (see Eq.~\eqref{1c} with the parameters of section \ref{II}), $\rho^{vib}_2$ appears to be negligible.  
   \label{fig:3}}
\end{figure}

In fact, one may even be more restrictive than that for the following reasons. First, $\rho^{tr}_{n'}(R_f,P_f)$ forces the kinetic energy to be equal to $E-E_{n'}$, as deduced from Eqs.~\eqref{5} and~\eqref{BB.11}. Second, $\rho^{vib}_{n'}(r_f,p_f)$ turns out to be negligible for states $(r_f,p_f)$ corresponding to vibrational energies larger than $\sim\;E_{n'}+2.5$, as illustrated in Fig.~\ref{fig:3} for $n'=2$. Consequently, the product $\rho^{tr}_{n'}(R_f,P_f)\; \rho^{vib}_{n'}(r_f,p_f)$ in Eq.~\eqref{23} limits the integration to phase space states belonging to the energy range [$E-E_{n'}$,$E+2.5$]. 
Defining $f(H)$ by 1 for $H$ within the previous range, and 0 for $H$ outside it,
Eq.~\eqref{23} can then be rewritten as
\\
\begin{equation}
|Q_{n'n}|^2 = (2\pi \hbar)^2 
\int\;dR_0dr_0dP_0dp_0\; \rho_{0}(R_0,r_0,P_0,p_0)\;f(H)
\; \rho^{tr}_{n'}(R_f,P_f)\; \rho^{vib}_{n'}(r_f,p_f).
\label{24bis}
\end{equation}
\\
The validity of Eq.~\eqref{23} is thus conditioned by the ability of classical mechanics to propagate $\rho_{t}(R,r,P,p)f(H)$ in a realistic way.

\begin{figure}
  \includegraphics[width=180mm]{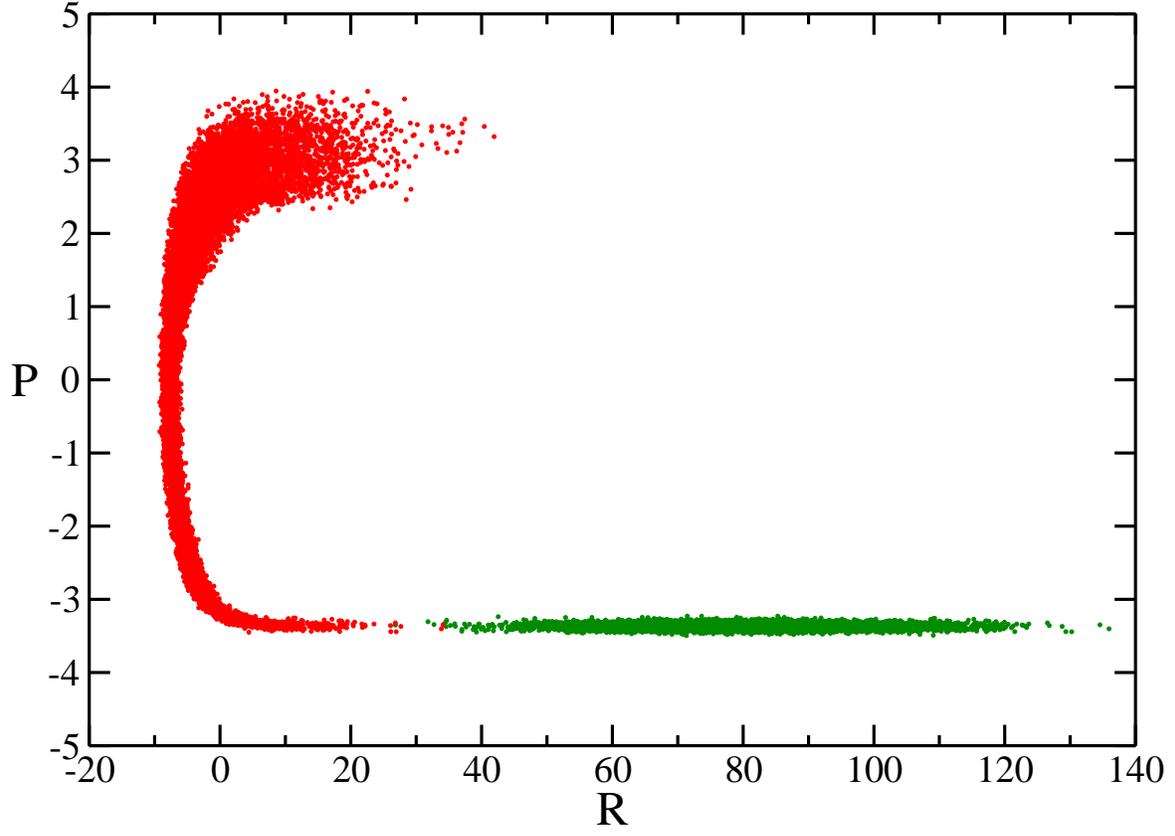}
  \caption{Green cloud: projection of $\rho_{0}(R,r,P,p)f(H)$ onto the $(R,P)$ plane for 
$n'=2$, $\epsilon = 0.05$ and $k_i = k_n$. Red cloud: projection of the classically propagated previous cloud when it bounces against the repulsive wall of the Secrest-Johnson PES.   
   \label{fig:4}}
\end{figure}

The projection of $\rho_{0}(R,r,P,p)f(H)$ onto the $(R,P)$ plane is represented by the green cloud in Fig.~\ref{fig:4} for 
$n'=2$, $\epsilon = 0.05$ and $k_i = k_n$ (see Eq.~\eqref{5}). The whole density lies beyond $R = R_f = 20$ units, defining the frontier of the interaction region for the Secrest-Johnson potential energy surface (PES) \cite{Secrest} (see Eq.~\eqref{1b}). The projection of the classically propagated density when it bounces against the repulsive wall of the PES is given by the red cloud in the same graphic. At this instant, the compression of the wave-packet is maximum.

Due to the small value of $\epsilon$, the green cloud is very narrow along the $P$ direction, and very broad along the $R$ direction, by virtue of the uncertainty principle. The red cloud extends from -10 to 20 units mainly along the $R$-axis, which turns out to be the whole interaction region available at the energy $E=10$. 

\begin{figure}
  \includegraphics[width=180mm]{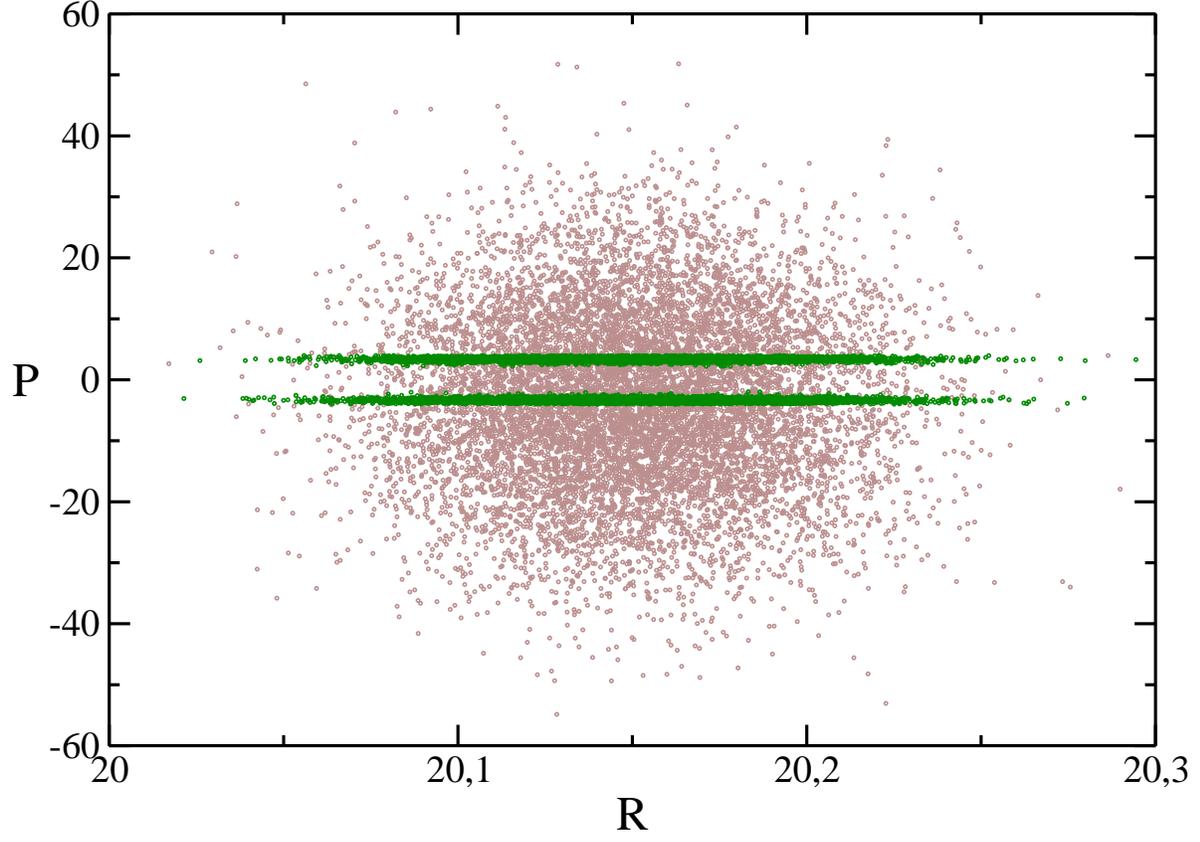}
  \caption{Brown cloud: projection of $\rho_{0}(R,r,P,p)$ onto the $(R,P)$ plane for 
$n'=2$, $\epsilon = 20$ and $k_i = k_n$. Green cloud: projection of $\rho_{0}(R,r,P,p)f(H)$ for the same parameters.
   \label{fig:5}}
\end{figure}

The projections of $\rho_{0}(R,r,P,p)$ and $\rho_{0}(R,r,P,p)f(H)$ are represented in Fig.~\ref{fig:5} for $\epsilon = 20$,
a much larger value than previously, the remaining parameters being unchanged. The first projection corresponds to the brown cloud while the second corresponds
to the two green ones. These clouds are now very broad along the $P$ direction, 
and very narrow along the $R$ direction, at the opposite of the previous case.
As a matter of fact, multiplying $\rho_{0}(R,r,P,p)$ by $f(H)$ strongly alters it for large values of $\epsilon$. 

On the other hand, the same operation has not effect for $\epsilon = 0.05$, i.e., projecting $\rho_{0}(R,r,P,p)$ onto the $(R,P)$ plane still leads to the green cloud in Fig.~\ref{fig:4}.

\begin{figure}
  \includegraphics[width=180mm]{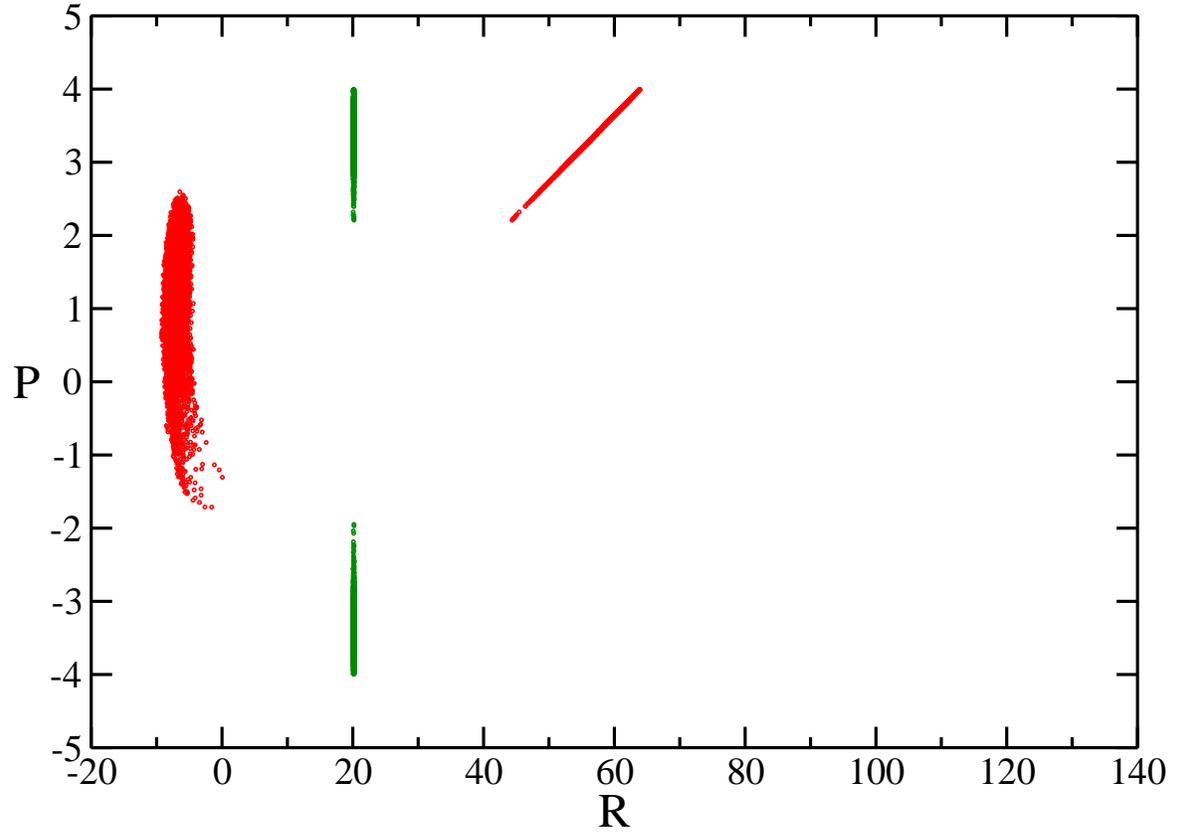}
  \caption{Same as Fig.~\ref{fig:4} for $n'=2$, $\epsilon = 20$ and $k_i = k_n$.   
   \label{fig:6}}
\end{figure}

Fig.~\ref{fig:6} is analogous to Fig.~\ref{fig:4} for $\epsilon = 20$. The left and right red clouds come from the classical propagation of the lower and upper green clouds, respectively. 

For $n'$ different from $n$, the right red cloud will not contribute to $|Q_{n'n}|^2$. This cloud is indeed
obtained by classically propagating $\rho_{0}(R,r,P,p)f(H)\Theta(P)$. Now, 
the previous density contains the factorized term $\rho^{vib}_{n}(r,p)$ (see Eq.~\eqref{22}) which remains
unaltered by the outward propagation and is orthogonal to $\rho^{vib}_{n'}(r,p)$. The overlap between these two
densities in Eq.~\eqref{24bis} is thus zero 
(note that the classical propagation is exact here, the PES being flat along the $R$-axis and quadratic along the
$r$-axis, satisfying thereby the conditions for which the Wigner equation reduces to the Liouville one \cite{SC6a,Wig,LS}). 
$|Q_{n'n}|^2$ is thus only due to the left red cloud, which extends from about -9 to -4 units along the $R$-axis. 

\begin{figure}
  \includegraphics[width=180mm]{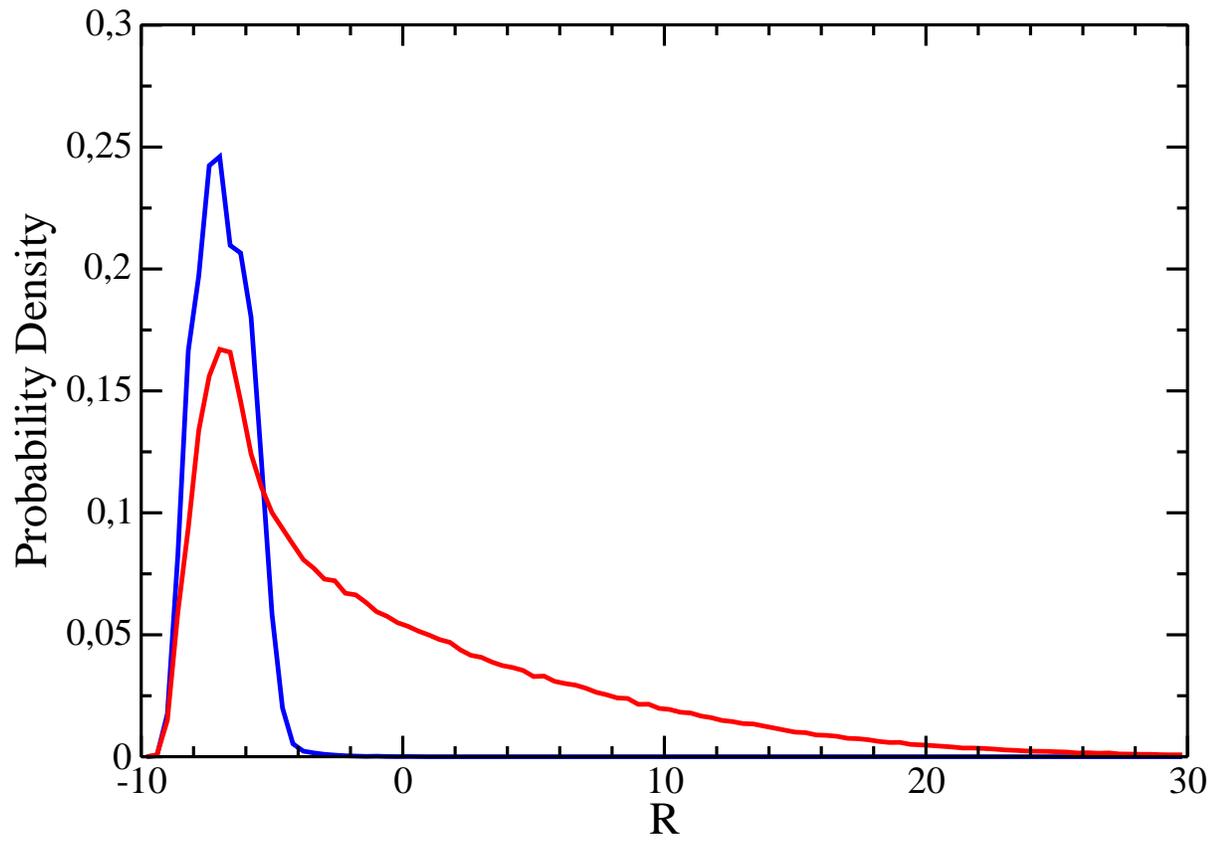}
  \caption{Distributions of $R$ obtained from the red cloud in Fig.~\ref{fig:4} (red curve, corresponding to $\epsilon = 0.05$) and the inner red cloud in Fig.~\ref{fig:6} (blue curve corresponding to $\epsilon = 20$). 
   \label{fig:7}}
\end{figure}

The distributions of $R$ obtained from the bouncing distributions are represented in Fig.~\ref{fig:7}. As anticipated from
the shape of the red cloud in Fig.~\ref{fig:4} and the inner red cloud in Fig.~\ref{fig:6}, 
the spreading of the distribution is comparable to the size of the interaction region
for $\epsilon = 0.05$, while it is $\sim 6$ times smaller than the same region for $\epsilon = 20$. 
Now, it is well known \cite{SC3,SC6a} that
the smaller the extention of a wave-packet in the configuration space, the more accurate the classical propagation of its
corresponding Wigner density. The basic reason is that in this limit, one may reasonably approximate the PES by a second order development around the wave-packet, which reduces the Wigner equation to the Liouville one \cite{SC3,SC6a,LS,Wig}.
As a consequence, large values of $\epsilon$, which strongly narrow the $R$-extention of $\rho_{t}(R,r,P,p)f(H)$ within
the interaction region, are expected to make its classical propagation more realistic, and the prediction of $|Q_{n'n}|^2$
more accurate than small values of $\epsilon$. 

On the other hand, there is no control of the $r$-extention of $\rho_{t}(R,r,P,p)f(H)$ in the present approach. This
makes it better suited to collisions where the reagents are prepared in the lowest excited vibrational states, less
spreaded along the $r$ coordinate.

For $n'$ equal to $n$, Eq.~\eqref{23} has less chance to be valid in the limit of large $\epsilon$, as the classically propagated density $\rho_{t}(R,r,P,p)$ cannot take into account the possible interference between the outgoing and incoming parts of the initial wave-packet, corresponding to the upper and lower green clouds in Fig.~\ref{fig:6}. 
However, setting
\
\begin{equation}
P_{n'n} = |S_{n'n}|^2,
\label{25a}
\end{equation}
$|Q_{n'n}|^2$ is found from Eq.~\eqref{9} to satisfy 
\\
\begin{equation}
g(k_n)^2 P_{n'n}+g(-k_n)^2 \delta_{n'n}+I = 
\left[\frac{\hbar^2 k_n}{\mu}\right] |Q_{n'n}|^2,
\label{25new}
\end{equation}
\\
where $I$ results from the interference between the first and second terms on the left-hand-side of Eq.~\eqref{9}.
This interference is of the same nature as the previous one, but expressed in a time-independent framework. 
Since Eq.~\eqref{23} cannot deal with it, it makes sense to remove $I$ from Eq.~\eqref{25new}, thus assuming 
that the population $P_{n'n}$ satisfies
\\
\begin{equation}
g(k_n)^2 P_{n'n}+g(-k_n)^2 \delta_{n'n} = 
\left[\frac{\hbar^2 k_n}{\mu}\right] |Q_{n'n}|^2.
\label{26new}
\end{equation}
\\
An analytical example is considered in Appendix B which supports the previous assumption. 
Note that when $n'$ is different from $n$, Eq.~\eqref{26new} is exact.

The main finding of the present analysis is that $\epsilon$ in Eq.~\eqref{22} has to be taken at a large value
in order to maximize the accuracy of the semiclassical Wigner method.

\subsection{Backward description of the dynamics}
\label{IV.E}

At the boundary of the interaction region, i.e., at $R=R_f$, Eq.~\eqref{1} becomes
\\
\begin{equation}
H=\frac{P^2_f}{2\mu}+\frac{p^2_f}{2m}+\frac{1}{2}m\omega^2(r_f-r_e)^2.
\label{23new}
\end{equation}
\\
Using Eqs.~\eqref{5},~\eqref{BB.11} and~\eqref{23new}, Eq.~\eqref{23} can be rewritten as
\\
\begin{equation}
|Q_{n'n}|^2 = 2\pi \hbar
\int\;dR_0dr_0dP_0dp_0\; \rho_{0}(R_0,r_0,P_0,p_0)
\; \rho^{vib}_{n'}(r_f,p_f)\; \delta\left[H-\frac{p^2_f}{2m}-\frac{1}{2}m\omega^2(r_f-r_e)^2-E+E_{n'}\right].
\label{24new}
\end{equation}
\\
$P_f$ being necessarily positive, the term $\Theta(P_f)$, coming from Eq.~\eqref{BB.11}, is not necessary to specify 
in the above integral. In refs.~\cite{SC10,Arbelon,Back}, the alternative set of coordinates $(t,H,r_f,p_f)$ was used in place of $(R_0,r_0,P_0,p_0)$. 
In these new coordinates, the origin of time corresponds to the instant where the system is at $R_f$.
The pair $(r_f,p_f)$ specifies the internal state of BC at time 0 and $H$ forces 
$P_f$ to take the value 
\\
\begin{equation}
P_f=\left[2\mu\left(H-\frac{p^2_f}{2m}-\frac{1}{2}m\omega^2(r_f-r_e)^2\right)\right]^{1/2}
\label{25}
\end{equation}
\\
(see Eq.~\eqref{23new}). $(R_f,P_f,r_f,p_f)$ lies along a given trajectory.
Now, any point along this trajectory can be reached from the previous point by moving along the trajectory  
a given period of time $|t|$ either forward ($t > 0$) or backward ($t < 0$). In other words, given $R_f$,
$(H,r_f,p_f)$ imposes the classical path, and $t$ the location along it. 
Consequently, $(t,H,r_f,p_f)$ allows to span the whole phase space. 

In addition to that, one may show the important property \cite{SC10,Back,Gutz,Kay1,Kay2}
\\
\begin{equation}
dR_0dr_0dP_0dp_0=dt dH dr_f dp_f
\label{26}
\end{equation}
\\
(see, in particular, Appendix C of ref.~\cite{SC10}). 
From Eqs.~\eqref{24new} and \eqref{26} and the straightforward integration 
with respect to $H$, we finally arrive at
\begin{equation}
|Q_{n'n}|^2 = 2\pi \hbar
\int\;dr_f dp_f\; \rho^{vib}_{n'}(r_f,p_f)\int_{-\infty}^{+\infty}\;dt\; \rho_{0}(R_t,r_t,P_t,p_t),
\label{27}
\end{equation}
\\
not forgetting that the delta term in Eq.~\eqref{BB.11} imposes the condition
\\
\begin{equation}
P_f=\hbar k_{n'}.
\label{28}
\end{equation}
\\
Finally, $(R_t,r_t,P_t,p_t)$ in Eq.~\eqref{27} is the value of $(R,r,P,p)$ at time $t$ when starting from 
$(R_f,r_f,P_f,p_f)$ at time 0 (the meaning of $(R_t,r_t,P_t,p_t)$ is thus different here and in section~\ref{IV.C}).

To summarize, the internal state $(r_f,p_f)$ of BC is randomly chosen within appropriate boundaries.
Together with Eq.~\eqref{28}, they allow to generate a trajectory from $R_f$ at time 0. The trajectory is then propagated
backward in time, i.e., in the direction of the reagent molecule, and $\rho_0(R_t,r_t,P_t,p_t)$ 
is time-integrated until the trajectory gets back to the separated fragments and $\rho_0(R_t,r_t,P_t,p_t)$ is found to be
negligible (in principle, one should also propagate 
forward in time. This is useless, however, as $\rho_0(R_t,r_t,P_t,p_t)$ is zero for positive values of $t$). 
The result is multiplied by the statistical weight $\rho^{vib}_{n'}(r_f,p_f)$
in order to get the integrand of Eq.~\eqref{27}. 

Eq.~\eqref{27} is, however, not in its final form. It can indeed be analytically simplified by integrating over $t$ as follows. 
$\rho_{0}(R_t,r_t,P_t,p_t)$ is given by Eq.~\eqref{22} where we assume that $\epsilon$ has a large value so as to maximize
the accuracy of $|Q_{n'n}|^2$, as seen in the previous section. $\rho_{0}(R,r,P,p)$ is thus narrow along the $R$-axis, 
stretched along the $P$-axis, and extends along the latter in the upper and lower half planes, as illustrated in 
Figs.~\ref{fig:5} and~\ref{fig:6}. Along the trajectory defined by ($R_f,r_f,P_f=\hbar k_{n'},p_f$), we have
\\
\begin{equation}
R_t=R_f+\frac{P_f}{\mu}t
\label{30}
\end{equation}
\\
within the phase space region corresponding to the free fragments with $P$ positive (after the collision).
Moreover,
\\
\begin{equation}
R_t=R_i+\frac{P_i}{\mu}[t-t(r_f,p_f)]
\label{30a}
\end{equation}
\\
within the phase space region corresponding to the free fragments with $P$ negative (before the collision).
Both $P_i$ and $t(r_f,p_f)$ depend on the ``initial conditions'' $(r_f,p_f)$ within the backward picture
of the dynamics. Note that $P_i$ has a negative value, since it corresponds to A approaching from BC.
  
From the previous considerations, the fact that $\rho_{0}(R,r,P,p)$ lies in the phase space part associated with 
the free fragments, and Eq.~\eqref{22}, we have
\\
\begin{equation}
\int_{-\infty}^{+\infty}\;dt\; \rho_{0}(R_t,r_t,P_t,p_t)=Q_{in}+Q_{out}
\label{31}
\end{equation}
with
\begin{equation}
Q_{in}=\frac{1}{\pi\hbar}\; \rho^{vib}_{n}(r_i,p_i)
e^{-\left(\frac{P_i+\hbar k_i}{\hbar \epsilon}\right)^2} 
\int_{-\infty}^{+\infty}\;dt\; e^{-\epsilon^2\left(\frac{P_i}{\mu}[t-t(r_f,p_f)]\right)^2}
\label{31a}
\end{equation}
and
\begin{equation}
Q_{out}=\frac{1}{\pi\hbar}\; \rho^{vib}_{n}(r_f,p_f)
e^{-\left(\frac{P_f+\hbar k_i}{\hbar \epsilon}\right)^2} 
\int_{-\infty}^{+\infty}\;dt\; e^{-\epsilon^2\left(R_f-R_i+\frac{P_f}{\mu}t\right)^2}.
\label{31b}
\end{equation}
\\
Note that, \emph{stricto sensu}, the upper time limit in $Q_{in}$ and the lower time limit in $Q_{out}$ cannot be
$+\infty$ and $-\infty$, respectively. They should be intermediates times, the former being lower than the latter. 
However, the region where $\rho_{0}(R,r,P,p)$ takes significant values is 
expected to be crossed very quickly in both directions, justifying the passage to infinity. 
Also, since $\rho^{vib}_{n}(r_t,p_t)$ does only depend on the vibrational energy 
corresponding to $(r_t,p_t)$, and this energy is a constant of motion in the asymptotic channel, $r_t$ and $p_t$
have been replaced in $Q_{in}$ by $r_i$ and $p_i$, their values at time $t$ where $R_t=R_i$. Moreover,
they have been replaced in $Q_{out}$ by $r_f$ and $p_f$.

Using the fact that
\begin{equation}
\int_{-\infty}^{+\infty}\;dx\;\frac{e^{-\left(x/\epsilon\right)^2}}{\pi^{1/2}\epsilon}=1
\label{31c}
\end{equation}
\\
in order to analytically integrate $Q_{in}$ and $Q_{out}$, and using Eqs.~\eqref{3},~\eqref{26new} and~\eqref{31}, we arrive
at
\\
\begin{equation}
P_{n'n} = 2\pi \hbar
\int\;dr_f dp_f\; \rho^{vib}_{n'}(r_f,p_f)\rho^{vib}_{n}(r_i,p_i) U(P_i)
\label{32}
\end{equation}
with
\begin{equation}
U(P_i)=
\frac{\hbar k_n}{|P_i|}\;
exp\left[\left(\frac{\hbar k_n-\hbar k_i}{\hbar \epsilon}\right)^2-
\left(\frac{P_i+\hbar k_i}{\hbar \epsilon}\right)^2\right].
\label{32a}
\end{equation}
\\
In addition, keeping $\epsilon$ at a large value so as to maximize the accuracy of the method clearly reduces
$U(P_i)$ to $\hbar k_n/|P_i|$. We finally obtain the first key expression of this work:
\\
\begin{equation}
P_{n'n} = 2\pi \hbar
\int\;dr_f dp_f\; \rho^{vib}_{n'}(r_f,p_f)\rho^{vib}_{n}(r_i,p_i) \frac{\hbar k_n}{|P_i|}.
\label{32b}
\end{equation}
\\

\begin{figure}
  \includegraphics[width=180mm]{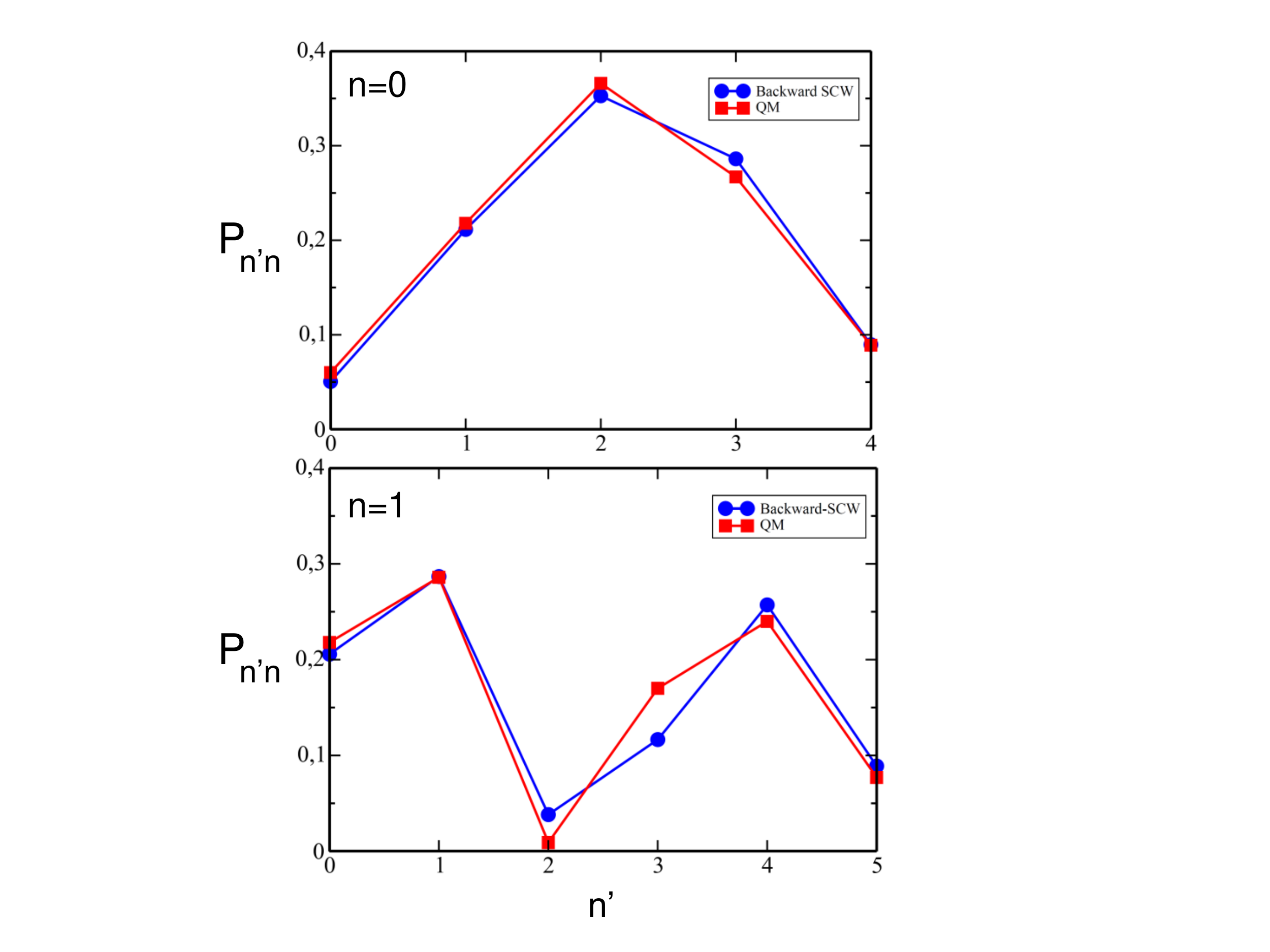}
  \caption{Final vibrational state distribution for the Secrest-Johnson model
  of collinear collision He+H$_2(n)$ $\longrightarrow$ He+H$_2(n')$ at a total energy of 10. 
  Backward-SCW predictions, given by Eq.~\eqref{32b}, are compared with QM results.   
   \label{fig:8}}
\end{figure}

In the case of the Secrest-Johnson model of He+H$_2$ collinear collision
\cite{Secrest}, Eq.~\eqref{32b} leads to the blue curves in Fig.~\ref{fig:8} (Backward-SCW results). 
A simple Monte-Carlo approach based on 10$^5$ trajectories for each value of $n'$
has been used.
The agreement with the QM results is very satisfying. The norm, i.e., the sum of
final state populations, appears to be very close to 1 for both $n=0$ and $n=1$.

\subsection{Use of microreversibility to reconnect with the standard forward description}
\label{IV.F}

Eq.~\eqref{32b} is actually not the expression most suitable for numerical applications. In order to get $P_{n'n}$, one
has to run a batch a trajectories backward in time from $R_f$ with $P_f$ given by Eq.~\eqref{28}. If $n'$ runs
from 0 to $n'_{max}$, $n'_{max}+1$ batches of trajectories must be run in order to obtain the whole final
state distribution. 

However, the microreversibility principle tells us that, within the framework of quantum mechanics, $P_{n'n}=P_{nn'}$. 
Hence, if the semiclassical Wigner
treatment is realistic, one may rewrite $P_{n'n}$ as  
\\
\begin{equation}
P_{n'n} = 2\pi \hbar
\int\;dr_f dp_f\; \rho^{vib}_{n}(r_f,p_f)\rho^{vib}_{n'}(r_i,p_i)
\frac{\hbar k_{n'}}{|P_i|}.
\label{33}
\end{equation}
\\
Eq.~\eqref{33} is just Eq.~\eqref{32b} with $n$ and $n'$ interchanged. $P_f$ is now given by
\\
\begin{equation}
P_f=\hbar k_{n}.
\label{34}
\end{equation}
\\
Last but not least, one may arbitrarily interchange the indices $i$ and $f$ in Eq.~\eqref{33},
i.e., call the initial internal conditions $r_i$ and $p_i$, keep the radial momentum at the negative value
\\
\begin{equation}
P_i=-\hbar k_{n},
\label{35}
\end{equation}
\\
run from $R_f$ the resulting trajectories forward in time, and call the final conditions $r_f$, $p_f$ and $P_f$
once the trajectory is back to $R_f$.
The ensemble of trajectories run is exactly the same as the one involved in Eq.~\eqref{33}. The only difference
is now that $P_f$ is positive, contrary to $P_i$ in Eq.~\eqref{33}. We finally get the second key expression of 
this work:
\\
\begin{equation}
P_{n'n} = 2\pi \hbar
\int\;dr_i dp_i\; \rho^{vib}_{n}(r_i,p_i)\rho^{vib}_{n'}(r_f,p_f)\frac{\hbar k_{n'}}{P_f}.
\label{36}
\end{equation}
\\
To sum up, the whole final state populations are determined from only one batch of trajectories run from $R_f$, 
with $(r_i,p_i)$ randomly chosen within appropriate boundaries and $P_i$ given by Eq.~\eqref{35}. Moreover, 
the indices $i$ and $f$ refer to initial and final conditions, respectively, and trajectories are integrated
forward in time. We have thus reconnected with the traditional way of simulating a molecular collision. 

\begin{figure}
  \includegraphics[width=180mm]{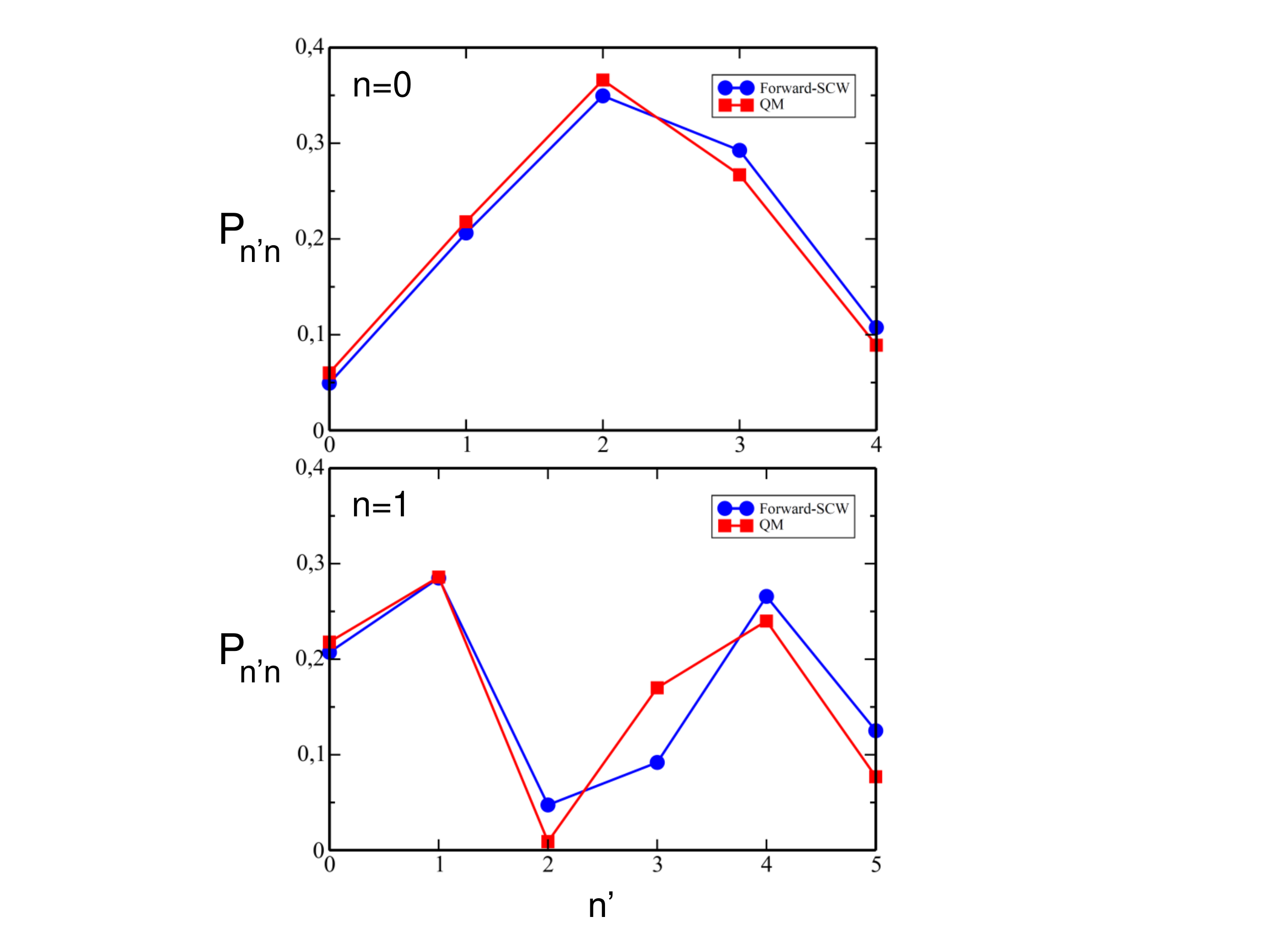}
  \caption{Final vibrational state distribution for the Secrest-Johnson model
  of collinear collision He+H$_2(n)$ $\longrightarrow$ He+H$_2(n')$ at a total energy of 10. 
  Forward-SCW predictions, given by Eq.~\eqref{36}, are compared with QM results. 
   \label{fig:9}}
\end{figure}

Eq.~\eqref{36} leads to the blue curves in Fig.~\ref{fig:9} (Forward-SCW results). They have been obtained from 10$^6$ trajectories, with $r_i$ and $p_i$ both randomly selected within the range [-6,6]. 
The agreement with the QM results is still very satisfying for $n=0$, a bit less for $n=1$ as compared with
the Backward method. This illustrates the fact that microreversibility is not strictly satisfied by the semiclassical 
Wigner method. The norm is again very close to 1 for both $n=0$ and $n=1$. One notes that the 
Forward-SCW results are in excellent agreement with the Lee-Scully ones (see Figs.~\ref{fig:1} and \ref{fig:2}).

\subsection{Recovering the expression of Lee and Scully}
\label{IV.G}

\begin{figure}
  \includegraphics[width=180mm]{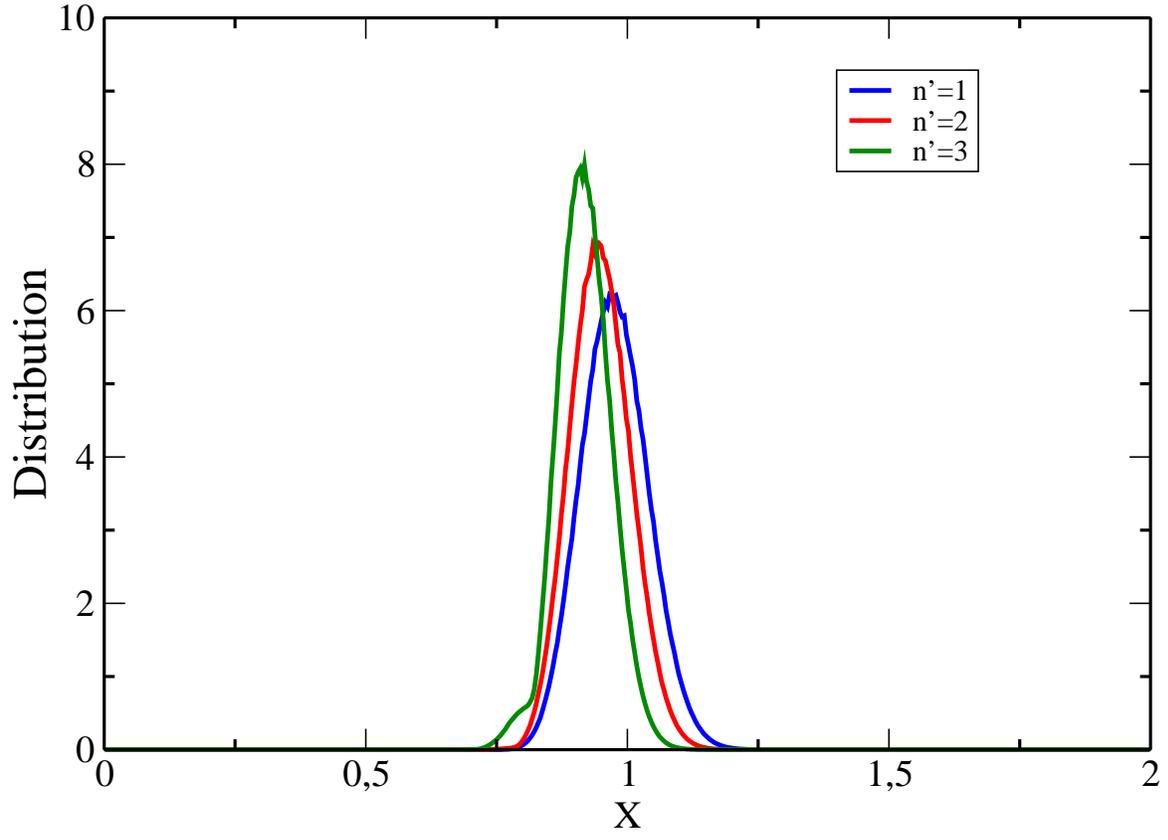}
  \caption{Distribution of the quantity $X=\hbar k_{n'}/P_f$ for $n=0$ and $n'=1-3$. 
   \label{fig:10}}
\end{figure}

The distribution of the ratio $\hbar k_{n'}/P_f$, weighted by $|\rho^{vib}_{n}(r_i,p_i)\rho^{vib}_{n'}(r_f,p_f)|$ in order to 
take into account the values which more contribute to $P_{n'n}$ (see Eq.~\eqref{36}), 
is represented in Fig.~\ref{fig:10} for $n=0$ and $n'=1-3$, corresponding
to the three largest populations (see the upper panel in Fig.~\ref{fig:1}). Similar distributions are 
found for $n=1$.
They appear to be peaked around 1. One may thus approximate $\hbar k_{n'}/P_f$ by 1 in 
Eq.~\eqref{36}, hence, recovering Eq.~\eqref{2B11}.

The reason why $\hbar k_{n'}/P_f$ is on average close to 1 in the model of He+H$_2$ collision considered
here is that a large part of the available energy $E$ is channeled into the recoil motion. In the limit where the
vibrational energy $E_{n'}$ is negligible as compared to $E$, Eq.~\eqref{5} leads to
\begin{equation}
\hbar k_{n'} \approx (2\mu E)^{1/2}.
\label{39}
\end{equation}
Moreover, 
\begin{equation}
H = \frac{\hbar^2 k_n^2}{2\mu}+\frac{p_i^2}{2m}+\frac{1}{2}m\omega^2(r_i-r_e)^2,
\label{40}
\end{equation}
\\
with $k_n$ given by Eq.~\eqref{5}. Hence, we have from Eqs.~\eqref{25} and~\eqref{40}
\\
\begin{equation}
P_f=\left[2\mu\left(E-E_n+\frac{p_i^2}{2m}+\frac{1}{2}m\omega^2(r_i-r_e)^2
-\frac{p^2_f}{2m}-\frac{1}{2}m\omega^2(r_f-r_e)^2\right)\right]^{1/2}.
\label{41}
\end{equation}
\\
But if $E$ is much larger than the vibrational energies involved in the problem, we also have
\\
\begin{equation}
P_f \approx (2\mu E)^{1/2}.
\label{42}
\end{equation}
\\
Since both $\hbar k_{n'}$ and $P_f$ have nearly the same value, their ratio
is roughly equal to 1. This explains the shape of the distributions displayed in
Fig.~\ref{fig:10}. 

However, this reasoning is no longer valid when $E_{n'}$ may be larger than the recoil energy. 
In such a case, there is \emph{a priori} no reason for replacing $\hbar k_{n'}/P_f$ by 1 in Eq.~\eqref{36}.
The same reasoning holds for $\hbar k_{n'}/|P_i|$ in Eq.~\eqref{32b}.

Last but not least, the Forward-SCW and Lee-Scully norms obtained from 5.10$^6$ trajectories (with $r_i$ and $p_i$ still randomly selected within the range [-6,6]) have been calculated. For $n=0$, the norms obtained by means of Eq.~\eqref{36}(Eq.~\eqref{2B11}) are 0.9984(0.9968), 0.9993(0.9985), 0.9995(0.9984) and 1.0003(0.9995) for $E=$ 10, 8, 6 and 3, respectively. For $n=1$, the analogous numbers are 0.9952(0.9922), 0.9907(0.9881), 0.9918(0.9862) and 1.0002(0.9980). Moreover, it has been checked that the number a trajectories run is sufficiently large for guarantying the order of the norms between the two approaches. As can be observed, the norm seems to be systematically more underestimated by the Lee-Scully expression than 
by the Forward-SCW one. Though the differences between the two sets of results
are very small, they nevertheless support the idea that the Lee-Scully expression is an approximation to the 
Forward-SCW expression (though an excellent one). 
Note that for $n=2$, the norms are on average equal to $\sim$ 0.94 and the previous order is lost. However,
the extention of the initial phase space density along the $r$ coordinate (see Eq.~\eqref{22}) 
is larger than for $n=0$ and 1, thus making the semiclassical Wigner predictions less reliable.

\section{Conclusion}
\label{V}

About three decades ago, Lee and Scully proposed a semiclassical Wigner treatment of bimolecular collisions \cite{SC5a}
leading to final state populations in surprisingly good agreement with exact quantum ones in the case of the
Secrest-Johnson model of collinear inelastic collision between He and H$_2$  \cite{Secrest}. 

The aim of the present paper was to provide a derivation from first principles of the previous method, following
the quantum description as far as possible. 
The derivation combines elements of (i) the time-dependent approach to semiclassical dynamics \cite{SC3}, (ii) 
the semiclassical Wigner treatment of photodissociation dynamics \cite{SC6a,SC6}, the backward description of 
molecular collisions \cite{SC10,Back} and (iv) microreversibility. 

The Lee-Scully expression of final state populations (Eq.~\eqref{2B11}) turns out to be the limiting form 
of a more general expression (Eq.~\eqref{36}) when a significant part of the product available energy is channeled into
the recoil motion. 

The present derivation is focused on collinear inelastic collisions, but defines
the main lines of a strategy that could be applicable to realistic three-dimensional bimolecular processes.
\newpage

\renewcommand{\theequation}{A.\arabic{equation}}

\setcounter{equation}{0}

\section*{Appendix A: Detailed derivation of the link between transition amplitudes and 
time-evolved wave-packet}

\subsubsection{Projection $C_j(E)$ of $\phi_E^j(R,r)$ on $\Psi_0(R,r)$}
\label{A.1}

From Eqs.~\eqref{2},~\eqref{4} and~\eqref{6} and integration over $r$, we arrive at
\\
\begin{multline}
C_j(E) = \left[\frac{\mu}{\hbar^2 k_j}\right]^{1/2}\delta_{nj}\int\;dk\;g(k)e^{ikR_i}\frac{1}{2\pi}\int\;dR\;e^{i(k_j-k) R}
\\
+\left[\frac{\mu}{\hbar^2 k_n}\right]^{1/2}S_{nj}{}^*\int\;dk\;g(k)e^{ikR_i}\frac{1}{2\pi}\int\;dR\;e^{-i(k_n+k) R}.
\label{A1}
\end{multline}
\\
Since
\begin{equation}
\frac{1}{2\pi}\int\;dR\;e^{i(k_n-k) R} = \delta(k_n-k),
\label{A2}
\end{equation}
\\
we get, by integration over $k$,
\begin{equation}
C_j(E) = \left[\frac{\mu}{\hbar^2 k_n}\right]^{1/2}\left[\delta_{nj}\;g(k_n)\;e^{ik_nR_i}
+S_{nj}{}^*g(-k_n)\;e^{-ik_nR_i}\right].
\label{A3}
\end{equation}

\subsubsection{Scattered wave-packet}
\label{A.2}

For $t$ tending to infinity, the time-evolved wave-packet $\Psi_t(R,r)$ (see Eq.~\eqref{7}) entirely lies 
in the asymptotic channel and is moving outward.
Its overlap with the incoming part of $\phi_{E'}^n(R,r)$ (see Eq.~\eqref{4}) is thus zero and Eqs.~\eqref{4} and~\eqref{7}  
lead thus to
\\
\begin{equation}
\lim_{t \to +\infty}\Psi_t(R,r) = \sum_j\;\int\;dE'\;C_j(E')\sum_{n"}\;S_{n"j}\;\left[\frac{\mu}{2\pi\hbar^2 k_{n"}}\right]^{1/2}e^{ik_{n"}R}\chi_{n"}(r)e^{-iE't/\hbar}.
\label{A4}
\end{equation}

\subsubsection{Projection $Q_{n'n}$}
\label{A.3}

Replacing in the projection $Q_{n'n}$ (see Eq.~\eqref{8}) $\Psi_t(R,r)$ by the RHS of Eq.~\eqref{A4} and
integrating over $r$ leads to
\\
\begin{equation}
Q_{n'n} = \lim_{t \to +\infty} 
\sum_j\;\int\;dE'\;\frac{\mu}{\hbar^2}\left[\frac{1}{k_{n'}(E')k_{n'}(E)}\right]^{1/2}C_{j'}(E')S_{n'j}(E')
\frac{1}{2\pi}\int\;dR\;e^{i[k_{n'}(E')-k_{n'}(E)]R}e^{-iE't/\hbar}.
\label{A5}
\end{equation}
\\
For clarity's sake, the dependence of $k_{n'}$ and $S_{n'j}$ on the energy is explicitly specified in the
previous expression. Using Eq.~\eqref{A2}, Eq.~\eqref{A5} becomes 
\\
\begin{equation}
Q_{n'n} = \lim_{t \to +\infty} 
\sum_j\;\int\;dE'\;\frac{\mu}{\hbar^2}\left[\frac{1}{k_{n'}(E')k_{n'}(E)}\right]^{1/2}C_{j'}(E')S_{n'j}(E')
\delta[k_{n'}(E')-k_{n'}(E)]e^{-iE't/\hbar}.
\label{A6}
\end{equation}
Applying the theorem
\begin{equation}
\delta[f(x)] = \sum_k\;\frac{1}{|f'(x_k)|} \delta(x-x_k),
\label{A7}
\end{equation}
\\
where the $x_k$'s are solutions of $f(x)=0$ \cite{CDL}, we find 
\\
\begin{equation}
Q_{n'n} = \sum_j\;C_{j}(E)S_{n'j}e^{-iEt/\hbar}.
\label{A8}
\end{equation}
\\
From Eqs.~\eqref{A3} and~\eqref{A8}, we arrive at
\\
\begin{equation}
Q_{n'n} = \left[\frac{\mu}{\hbar^2 k_{n}}\right]^{1/2}
\left[g(k_n)\;e^{ik_nR_i}\;S_{n'n}+g(-k_n)\;e^{-ik_nR_i}\;\sum_j\;S_{nj}{}^*S_{n'j}\right]e^{-iEt/\hbar}.
\label{A9}
\end{equation}
\\
However, $\mathbf{S^\dagger S = 1}$, where $\mathbf{S}$ is the \emph{S}-matrix, thus implying
\\
\begin{equation}
\sum_j\;S_{nj}{}^*S_{n'j} = \delta_{n'n}.
\label{A10}
\end{equation}
\\
From Eqs.~\eqref{A9} and~\eqref{A10}, we finally arrive at Eq.~\eqref{9}.

\newpage

\renewcommand{\theequation}{B.\arabic{equation}}

\setcounter{equation}{0}

\section*{Analytical example supporting the use of Eq.~\eqref{26new}}

In this example, $k_i$ is kept at 0 in Eq.~\eqref{3}, so $g(k)=g(-k)$. Moreover, the potential energy is given by
\\
\begin{equation}
V(R,r)=\frac{1}{2} m \omega^2(r-r_e)^2
\label{B1}
\end{equation}
\\ 
for $R \ge$ 0, and $+\infty$ for $R < 0$. If so, no inelastic transition is possible and we have
\\
\begin{equation}
\phi_E^n(R,r) = \left[\frac{\mu}{2\pi\hbar^2 k_n}\right]^{1/2}\chi_n(r)\left[e^{-ik_n R}+S_{nn}e^{ik_{n}R}\right].
\label{B2}
\end{equation}
\\
Since $\phi_E^n(R,r)$ is 0 for $R=0$, 
\begin{equation}
S_{nn}=-1.
\label{B3}
\end{equation} 
Squaring both sides of Eq.~\eqref{9} leads to
\\
\begin{equation}
g(k_n)^2 \left[P_{nn}+1+2Re\left(S_{nn}e^{ik_{n}R_i}\right)\right] = 
\left[\frac{\hbar^2 k_n}{\mu}\right] |Q_{nn}|^2.
\label{B4}
\end{equation}
\\
This is Eq.~\eqref{25new} for $n'=n$. 

Besides, we may apply Eq.~\eqref{23}, together with Eqs.~\eqref{BB.11} and~\eqref{22}. In a first step, one can
replace $r_f$ and $p_f$ in $\rho^{vib}_{n}(r_f,p_f)$ by $r_0$ and $p_0$, owing to the fact that the vibrational energy 
$E_v$ is a constant of motion throughout the whole collision, and $\rho^{vib}_{n}(r_f,p_f)$ does only depend on $E_v$
(see Eqs.~\eqref{2B7}-\eqref{2B8a}). Since  
\\
\begin{equation}
2\pi\hbar \int\;dr_0dp_0\; \rho^{vib}_{n}(r_0,p_0){}^2 = 1
\label{B6}
\end{equation}
\cite{SC5a}, we arrive at
\begin{equation}
|Q_{nn}|^2 = 2 \int\;dR_0dP_0\; e^{-\epsilon^2(R_0-R_i)^2}
e^{-\left(\frac{P_0}{\hbar \epsilon}\right)^2}
\; \rho^{tr}_{n}(R_f,P_f).
\label{B7}
\end{equation}
\\
As the translational motion is unperturbed by the bouncing, $P_f$ in $\rho^{tr}_{n}(R_f,P_f)$ can be replaced 
by $P_0$. Using Eqs.~\eqref{BB.11},~\eqref{A7} and~\eqref{3}, one gets after integrating over $R_0$ and $P_0$,
\\
\begin{equation}
|Q_{nn}|^2=\frac{2\mu g(k_n)^2}{\hbar^2 k_n}.
\label{B8}
\end{equation}
\\
Since $P_{nn}=1$, Eq.~\eqref{B8} can be rewritten as
\\
\begin{equation}
g(k_n)^2 \left(P_{nn}+1\right) = 
\left[\frac{\hbar^2 k_n}{\mu}\right] |Q_{nn}|^2,
\label{B9}
\end{equation}
\\
which is just Eq.~\eqref{26new} for $n'=n$. The semiclassical expectation of $|Q_{n'n}|^2$ is thus solution
of Eq.~\eqref{B4}, or Eq.~\eqref{25new}, with the interference term removed.

\newpage

\newpage

\end{document}